\documentclass[preprintnumbers,article,amsmath,amssymb,floatfix,10pt,prd,twocolumn,
superscriptaddress,nofootinbib]{revtex4-2}
\usepackage{bm}
\usepackage{amsfonts}
\usepackage{latexsym}
\usepackage[latin1]{inputenc}
\usepackage{graphicx}
\usepackage{amsmath}
\usepackage{bigints} 
\usepackage{palatino}
\usepackage{mathpazo}
\usepackage{textcomp}
\usepackage{float}
\usepackage{booktabs}
\usepackage{dcolumn}
\usepackage{ragged2e}
\usepackage{hyperref}
\hypersetup{colorlinks,citecolor=blue}
\hypersetup{colorlinks=true,linkcolor=red,filecolor=magenta,    urlcolor=blue}
\usepackage{amsmath}
\usepackage{xcolor}
\usepackage{orcidlink}
\usepackage{epsfig}
\usepackage{subfigure}
\usepackage{commath}

\def\jnl@style{\it}
\def\aaref@jnl#1{{\jnl@style#1}}

\def\aaref@jnl#1{{\jnl@style#1}}

\def\aj{\aaref@jnl{AJ}}                   
\def\apj{\aaref@jnl{ApJ}}                 
\def\apjl{\aaref@jnl{ApJ}}                
\def\apjs{\aaref@jnl{ApJS}}               
\def\apss{\aaref@jnl{Ap\&SS}}             
\def\aap{\aaref@jnl{A\&A}}                
\def\aapr{\aaref@jnl{A\&A~Rev.}}          
\def\aaps{\aaref@jnl{A\&AS}}              
\def\mnras{\aaref@jnl{Mon.~Not.~Roy.~Astron.~Soc.}}             
\def\prd{\aaref@jnl{Phys.~Rev.~D}}        
\def\prc{\aaref@jnl{Phys.~Rev.~C}}  
\def\prl{\aaref@jnl{Phys.~Rev.~Lett.}}    
\def\qjras{\aaref@jnl{QJRAS}}             
\def\skytel{\aaref@jnl{S\&T}}             
\def\ssr{\aaref@jnl{Space~Sci.~Rev.}}     
\def\zap{\aaref@jnl{ZAp}}                 
\def\nat{\aaref@jnl{Nature}}              
\def\aplett{\aaref@jnl{Astrophys.~Lett.}} 
\def\apspr{\aaref@jnl{Astrophys.~Space~Phys.~Res.}} 
\def\physrep{\aaref@jnl{Phys.~Rep.}}      
\def\physscr{\aaref@jnl{Phys.~Scr}}       
\def\commat{\aaref@jnl{Comm.~Math.~Phys.}}              
\def\science{\aaref@jnl{Science}}               
\def\cqg{\aaref@jnl{Classical Quant.~Grav.}}            
\def\jpcs{\aaref@jnl{JPCS}}                                     
\def\ijmpd{\aaref@jnl{Int.~J.~Mod.~Phys.~D}}                    
\def\grg{\aaref@jnl{Gen.~Relat.~Gravit.}}               
\def\rpp{\aaref@jnl{Rep.~Prog.~Phys.}}          
\def\npa{\aaref@jnl{Nucl.~Phys.~A}}        
\def\lrr{\aaref@jnl{Living Rev.~Rel.}}                   
\def\jcap{\aaref@jnl{J.~Cosmology Astropart.~Phys.}}    
\def\rmp{\aaref@jnl{Rev.~Mod.~Phys.}}   
\def\epjc{\aaref@jnl{Eur.~Phys.~J.~C}} 
\def\plb{\aaref@jnl{~Phy.~Lett.~B}} 
\def\mpla{\aaref@jnl{Mod.~Phy.~Lett.~A}} 
\def\arxiv{\aaref@jnl{arxiv.org}}

\allowdisplaybreaks[1]

\begin{document}
\color{black} 
\title{Impact of dark matter galactic halo models on wormhole geometry within $f(Q,T)$ gravity}

\author{Moreshwar Tayde\orcidlink{0000-0002-3110-3411}}
\email{moreshwartayde@gmail.com}
\affiliation{Department of Mathematics, Birla Institute of Technology and
Science-Pilani,\\ Hyderabad Campus, Hyderabad-500078, India.}

\author{Zinnat Hassan\orcidlink{0000-0002-6608-2075}}
\email{zinnathassan980@gmail.com}
\affiliation{Department of Mathematics, Birla Institute of Technology and
Science-Pilani,\\ Hyderabad Campus, Hyderabad-500078, India.}

\author{P.K. Sahoo\orcidlink{0000-0003-2130-8832}}
\email{pksahoo@hyderabad.bits-pilani.ac.in}
\affiliation{Department of Mathematics, Birla Institute of Technology and
Science-Pilani,\\ Hyderabad Campus, Hyderabad-500078, India.}

%
\date{\today}

\begin{abstract}
This study investigates the possible existence of wormhole solutions with dark matter galactic halo profiles in the background of $f(Q,T)$ gravity. 
The primary focus of the current study is to find the significance of dark matter (DM) in the search for traversable wormhole solutions within galactic halos. Various dark matter profiles, such as Universal Rotation Curves (URC), Navarro-Frenk-White (NFW) model-I, and NFW model-II, are examined within two different $f(Q,T)$ models. The DM halo density profiles generate appropriate shape functions under the linear model that satisfy all the essential conditions for presenting the wormhole geometries. Apart from that, we take into account an embedded wormhole-specific shape function to inspect DM profiles under the non-linear model. We noticed that the null energy conditions are violated by the obtained solution from each model, which confirms that the DM support wormholes to sustain in the galactic halo. The findings reveal that the solutions obtained for different density profiles of dark matter halos within generalized symmetric teleparallel gravity demonstrate viability.
\end{abstract}

\maketitle


\section{Introduction}
The search for a theory of exotic objects over Einstein's general theory of relativity (GR) has acquired a significant amount of curiosity in the literature. A wormhole possesses one of the probable solutions to Einstein's field equations. Wormholes are links that connect two branes, universes, or even just two locations at the manifold. One of the earliest wormhole solutions, the Einstein-Rosen bridge, was discovered by Einstein and Rosen \cite{Einstein} in 1935. Initially, the Einstein-Rosen wormhole was found to be non-traversable, as its throat would rapidly expand and contract, preventing anything, not even a photon, from passing through \cite{J. A. Wheeler, K. Jusufi}. However, the issue of wormhole traversability was later addressed by Morris and Thorne \cite{M.S. Morris}. The Schwarzschild wormhole \cite{Dobrev} is the first form of wormhole solution present in the Schwarzschild metric defining an eternal black hole. Nevertheless, it turned out that it could have fallen too rapidly. In a nutshell, if there is an exotic sort of matter with negative energy density, the wormholes may be sustained.

In the last two decades, considerable research has been dedicated to exploring traversable wormhole solutions \cite{D. Hochberg2, F. S. N. Lobo, J. L. Blazquez-Salcedo, M. R. Mehdizadeh, E. Ayon-Beato, F. Parsaei, F. Canfora, G. Clement}. It is known that constructing wormhole solutions without \textit{exotic matter} (it is responsible for violating Null Energy Conditions (NEC) and is crucial for constructing a traversable wormhole) poses a significant challenge. To minimize the volume of exotic types of matter, Visser \cite{Visser1, Visser2} described a cut-and-paste method for constructing a spherically symmetric thin-shell wormhole where the exotic matter was located, allowing an observer to pass through unexpectedly without coming into contact with it. In \cite{Hayward}, the authors considered minimizing the violation of NEC for traversable wormholes and examined their stability. Moreover, Visser et al. \cite{Dadhich} claimed that it could reduce the violation of the NEC by choosing an appropriate geometry for the wormhole. It is known that Despite constructing wormhole solutions with ordinary matter being complicated, this problem has been addressed for rotating cylindrical wormholes in the framework of GR, and wormhole solutions preserving weak energy conditions (WEC) have been extensively explored in Refs. \cite{Bronnikov1, Bronnikov2}. Furthermore, it is worth noting that wormholes can also be supported by normal matter in modified theories of gravity \cite{Bhawal, Montelongo, SenGupta, Moraes1, Shaikh, R1, Kunz, Mak, Galiakhmetov, Moraes2, Sakalh, R10, M. Jamil, M. Sharif, Kavya 1, Vittorio 1, Vittorio 2, Vittorio 3}.\\
\indent In the late '90s, researchers introduced the concept of the non-metricity theory following the proposal of Symmetric Teleparallel Gravity \cite{Kalay, Nester, Conroy}. This theory, which is free from torsion and curvature, connects gravitation to the non-metricity tensor and the associated nonmetricity scalar. In 2018, Jimenez et al. developed a generalization of this theory known as $f(Q)$ gravity \cite{Jimenez}. In $f(Q)$ gravity, the gravitational field is described solely by the non-metricity scalar $Q$. This theory has been successful in explaining various perturbation and observational data, including Redshift Space Distortion, Supernovae type Ia, Cosmic Microwave Background, and Baryonic Acoustic Oscillations, etc., \cite{Koivisto11, Soudi, Salzano, Banos}. Notably, a study suggests that the non-metricity $f(Q)$ gravity may challenge the $\Lambda$CDM \cite{Anagnostopoulos}. The application of $f(Q)$ gravity extends to the astrophysical domain as well. Researchers have explored the behavior of black holes within the framework of $f(Q)$ gravity \cite{Fell}. Wormhole solutions \cite{Hassan11} as well as Casimir wormholes \cite{Ghosh} have been analyzed in $f(Q)$ gravity. In \cite{Zhai}, the spherically symmetric configurations in $f(Q)$ gravity have been discussed. Further interesting applications of $f(Q)$ gravity can be found in the literature \cite{Hohmann, Koivisto, Kuhn, Zhao, Wang2, Solanki 1, Calza, Sanjay 1, Gadbail 1}.\\
\indent In recent times, there has been growing interest in the field of $f(Q,T)$ theories, which explore a matter-geometry coupling represented by viable functions of the non-metricity scalar $Q$ and the trace of the energy-momentum tensor $T$ in the Lagrangian \cite{Y.Xu}. While this proposed gravity theory is relatively new, it holds considerable potential in various cosmological applications. Several studies have already discussed the cosmological implications of $f(Q, T)$ gravity. The initial works \cite{Y.Xu} covered the first cosmological significances of this theory, while later research \cite{Arora111} focused on investigating the late-time accelerated expansion with observational constraints for the $f(Q, T)$ gravity model. Additionally, various other aspects such as Baryogenesis \cite{Bhattacharjee}, Cosmological inflation \cite{Shiravand}, Reconstruction of the $f(Q, T)$ Lagrangian \cite{Gadbail 2}, and Cosmological perturbations \cite{Najera} have been widely explored. However, it is worth noting that limited attention has been given to studying the astrophysical implications of this modified gravity. In one work by Tayde et al. \cite{Tayde 1}, where they investigated static spherically symmetric wormhole solutions in $f(Q,T)$ gravity for both linear and non-linear models under different equations of state relations. Their findings revealed that exact solutions were attainable for the linear model, while the non-linear model posed considerable challenges in obtaining analytical solutions. Also, wormhole solutions in $f(Q,T)$ gravity using the MIT bag model has been studied \cite{Tayde 2}. Further, in \cite{Sneha 2}, the thin-shell gravatar model in $f(Q, T)$ gravity is discussed. recently, constant-roll and primordial black holes in $f(Q, T)$ gravity are analyzed in \cite{Bourakadi}. In addition, one can check more interesting articles on astrophysical works \cite{ Tayde 3, Kavya 2, Tayde 4}. Despite the growing interest in $f(Q, T)$ gravity, it remains a relatively nascent theory with untapped potential. Therefore, in light of the scarcity of research on the astrophysical aspect of this gravity theory, we were inspired to delve into investigating wormhole solutions with dark matter in the context of $f(Q,T)$ gravity.\\
\indent DM is a fundamental constituent of the Universe, detectable primarily through its gravitational effects rather than its luminosity. Approximately 97\% of the Universe's content comprises dark energy and DM, while only about 3\% consists of observable matter. The virial theorem was initially used by Zwicky \cite{F. Zwicky} to propose the existence of DM in galaxies. It is widely believed that spiral galaxies exhibit URC, and the presence of DM in their galactic halos can be verified by its gravitational influence on the URC \cite{M.S. Roberts, L. Roszkowski}. Based on observations, such as the NFW density profile and the flat rotation curves of galaxies, Rahaman et al. \cite{F. Rahaman} demonstrated that galactic halos could potentially support the existence of traversable wormholes. Consequently, the topic of galactic halo wormholes has been explored in various theories of gravity, including modified theories. For instance, Sharif et al. \cite{M. Sharif 1, M. Sharif 2, M. Sharif 3} investigated galactic halo wormhole solutions in different modified gravity theories. Mustafa et al. \cite{G. Mustafa 1} focuses on investigating the physical properties of DM in the galactic halo regime and wormhole geometry in the modified teleparallel gravity. Further, by examining the three distinct DM halo profiles, namely URC, NFW, and Scalar Field Dark Matter (SFDM), the potential existence of generalized wormhole geometry within the galactic halo regions was investigated \cite{G. Mustafa 2}. Another recent work focused on the evolution of topologically deformed wormholes in DM halos \cite{A. Ovgun}. Furthermore, the formation of spherically symmetric traversable wormholes in DM halos with isotropic pressure was explored by Xu et al. \cite{Z. Xu}. Additionally, Kuhfittig \cite{P. K. F. Kuhfittig} studied the phenomenon of gravitational lensing by galactic halo wormholes. In this study, we aim to investigate traversable wormhole solutions using DM halo profiles within the framework of the $f(Q,T)$ theory of gravity. To achieve this, we consider different DM density profile models, namely the URC model and the cold dark matter halo, with two different NFW models. This research aims to establish the existence of galactic halo traversable wormholes within the context of $f(Q, T)$ gravity for these considered models.\\
\indent The structure of this paper is outlined as follows: In Section \ref{sec2}, we provide an introduction to the fundamental formalism of $f(Q, T)$ gravity, considering the necessary traversability conditions and subsequently deriving the corresponding field equations. Section \ref{sec3} offers a concise overview of the linear form of $f(Q, T)$ as applied to DM halo profiles. In Section \ref{sec4}, we extend our review to encompass the non-linear form of $f(Q, T)$ concerning DM halo profiles. Finally, in the concluding section, we summarize and discuss the results obtained throughout this study.

\section{Traversability conditions of wormhole and field equations in $f(Q,T)$ gravity}
\label{sec2}
We begin by considering a spherically symmetric and static space-time described by the following metric:
\begin{equation} \label{1}
ds^{2}=e^{\nu(r)}dt^{2}-e^{\lambda(r)} dr^{2}-r^{2} d\theta^{2}-r^{2}\sin^{2}\theta d\Phi^{2}.
\end{equation}
Here, the metric components, $\nu(r) = 2\phi(r)$ and $e^{\lambda(r)} = \left(\frac{r - b(r)}{r}\right)^{-1}$, are functions that depend only on the radial coordinate. The shape of the wormholes is determined by the shape function $b(r)$. The function $\phi(r)$ represents the redshift function associated with gravitational redshift. The flaring-out condition must be satisfied by the shape function $b(r)$ for a wormhole to be traversable, expressed as $(b - b'r)/b^2 > 0$ \cite{M.S. Morris}. At the wormhole throat, denoted by $r_0$, the condition $b(r_0) = r_0$ is imposed, and $b^{\prime}(r_0) < 1$ ($b^{\prime}(r)$ denotes the derivative of $b(r)$ with respect to $r$). Moreover, the asymptotic flatness condition, which states that $\frac{b(r)}{r}\rightarrow 0$ as $r\rightarrow \infty$, is also required. Additionally, to avoid an event horizon, the function $\phi(r)$ must remain finite everywhere. In the context of Einstein's General Relativity, satisfying these criteria may indicate the existence of exotic matter at the wormhole throat.

Now, let us briefly overview some general aspects of $f(Q, T)$ gravity. In this context, we consider the action of symmetric teleparallel gravity, as proposed in the study \cite{Y.Xu}
\begin{equation}\label{1a}
\mathcal{S}=\int\frac{1}{16\pi}\,f(Q, T)\sqrt{-g}\,d^4x+\int \mathcal{L}_m\,\sqrt{-g}\,d^4x\, .
\end{equation}
The function involving the non-metricity scalar $Q$ and the trace of the energy-momentum tensor $T$ is denoted as $f(Q, T)$. Here, $g$ represents the determinant of the metric tensor $g_{\mu\nu}$, and $\mathcal{L}_m$ corresponds to the matter Lagrangian density.\\
The non-metricity tensor is defined by the following equation \cite{Jimenez}:
\begin{equation}\label{2}
Q_{\lambda\mu\nu}=\bigtriangledown_{\lambda} g_{\mu\nu}\,.
\end{equation}
Also, the superpotential or non-metricity conjugate can be defined as
\begin{equation}\label{3}
\hspace{-0.3cm} P^\alpha_{\,\,\,\mu\nu}=\frac{1}{4}\left[-Q^\alpha\;_{\mu\nu}+2Q_{(\mu}\;^\alpha\;_{\nu)}+Q^\alpha g_{\mu\nu}-\tilde{Q}^\alpha g_{\mu\nu}-\delta^\alpha_{(\mu}Q_{\nu)}\right].
\end{equation}
And two traces of the non-metricity tensor are given by
\begin{equation}
\label{4}
Q_{\alpha}=Q_{\alpha}\;^{\mu}\;_{\mu},\; \tilde{Q}_\alpha=Q^\mu\;_{\alpha\mu}.
\end{equation}
The non-metricity scalar is represented as \cite{Jimenez}
\begin{eqnarray}
\label{5}
Q &=& -g^{\mu\nu}\left(L^\beta_{\,\,\,\alpha\mu}\,L^\alpha_{\,\,\,\nu\beta}-L^\beta_{\,\,\,\alpha\beta}\,L^\alpha_{\,\,\,\mu\nu}\right),\\
&=& -Q_{\alpha\mu\nu}\,P^{\alpha\mu\nu},
\end{eqnarray}
 where $L^\beta_{\,\,\,\mu\nu}$ represents the disformation defined by 
\begin{equation}\label{6}
L^\beta_{\,\,\,\mu\nu}=\frac{1}{2}Q^\beta_{\,\,\,\mu\nu}-Q_{(\mu\,\,\,\,\,\,\nu)}^{\,\,\,\,\,\,\beta}.
\end{equation}

Now, the gravitational equations of motion can be obtained by varying the action with respect to the metric tensor $g_{\mu\nu}$ and are represented as
\begin{multline}\label{7}
\frac{-2}{\sqrt{-g}}\bigtriangledown_\alpha\left(\sqrt{-g}\,f_Q\,P^\alpha\;_{\mu\nu}\right)-\frac{1}{2}g_{\mu\nu}f + f_T \left(T_{\mu\nu} +\Theta_{\mu\nu}\right) \\
-f_Q\left(P_{\mu\alpha\beta}\,Q_\nu\;^{\alpha\beta}-2\,Q^
{\alpha\beta}\,\,_{\mu}\,P_{\alpha\beta\nu}\right)=8\pi T_{\mu\nu},
\end{multline}
where $f_Q=\frac{\partial f}{\partial Q}$ and $f_T=\frac{\partial f}{\partial T}$.

The energy-momentum tensor for the fluid depiction of space-time can be described as
\begin{equation}\label{8}
T_{\mu\nu}=-\frac{2}{\sqrt{-g}}\frac{\delta\left(\sqrt{-g}\,\mathcal{L}_m\right)}{\delta g^{\mu\nu}},
\end{equation}
and
\begin{equation}\label{9}
\Theta_{\mu\nu}=g^{\alpha\beta}\frac{\delta T_{\alpha\beta}}{\delta g^{\mu\nu}}.
\end{equation}
In this study, we conduct an analysis of wormhole solutions while considering an anisotropic energy-momentum tensor. The formulation of this tensor, as presented in \cite{M.S. Morris, Visser}, is represented by Eq. \eqref{11} as follows:
\begin{equation}\label{11}
T_{\mu}^{\nu}=\left(\rho+p_t\right)u_{\mu}\,u^{\nu}-p_t\,\delta_{\mu}^{\nu}+\left(p_r-p_t\right)v_{\mu}\,v^{\nu},
\end{equation}
Here, the variables involved are defined as follows: $\rho$ represents the energy density, $u_{\mu}$ and $v_{\mu}$ are the four-velocity vector and unitary space-like vectors, respectively. Both vectors satisfy the conditions $u_{\mu}u^{\nu} = -v_{\mu}v^{\nu} = 1$. Additionally, $p_r$ and $p_t$ denote the radial and tangential pressures, respectively, and both are functions of the radial coordinate $r$. The trace of the energy-momentum tensor is given by $T = \rho - p_r - 2p_t$.\\
In this article, we employ the matter Lagrangian $\mathcal{L}_m = -P$ \cite{Correa}, which leads to the following form for Eq. \eqref{9}:
\begin{equation}\label{12}
\Theta_{\mu\nu}=-g_{\mu\nu}\,P-2\,T_{\mu\nu},
\end{equation}
Here, $P$ represents the total pressure and can be expressed as $P = \frac{p_r + 2p_t}{3}$.
The non-metricity scalar $Q$ for the metric \eqref{1} is provided by \cite{Tayde 1} and is given by:
\begin{equation}\label{13}
Q=-\frac{b}{r^2}\left[\frac{rb^{'}-b}{r(r-b)}+2\phi^{'}\right].
\end{equation}\\
The corresponding field equations for $f(Q, T)$ gravity, in the context of this study, are presented as follows \cite{Tayde 1}:
\begin{multline}\label{14}
8 \pi  \rho =\frac{(r-b)}{2 r^3} \left[f_Q \left(\frac{b \left(2 r \phi '+2\right)}{r-b}+\frac{(2 r-b) \left(r b'-b\right)}{(r-b)^2}\right)
\right. \\ \left.
-\frac{2r^3 f_T (P+\rho )}{(r-b)}+\frac{2 b r f_{\text{QQ}} Q'}{r-b}+\frac{f r^3}{r-b}\right],
\end{multline}
\begin{multline}\label{15}
8 \pi  p_r=-\frac{(r-b)}{2 r^3} \left[f_Q \left(\frac{b }{r-b}\left(\frac{r b'-b}{r-b}+2 r \phi '+2\right)
\right.\right. \\ \left.\left.
-4 r \phi '\right)-\frac{2r^3 f_T \left(P-p_r\right)}{(r-b)}+\frac{2 b r f_{\text{QQ}} Q'}{r-b}+\frac{f r^3}{r-b}\right],
\end{multline}
\begin{multline}\label{16}
8 \pi  p_t=-\frac{(r-b)}{4 r^2} \left[f_Q \left(\frac{ \left(\frac{2 r}{r-b}+2 r \phi '\right) \left(r b'-b\right)}{r (r-b)}+
\right.\right. \\ \left.\left.
\frac{4 (2 b-r) \phi '}{r-b}-4 r \phi '' -4 r \left(\phi '\right)^2\right)
-4 r f_{\text{QQ}} Q' \phi '\right.\\\left.
-\frac{4r^2 f_T \left(P-p_t\right)}{(r-b)}+\frac{2 f r^2}{r-b}\right].
\end{multline}
Using these specific field equations, it becomes possible to explore different wormhole solutions within the framework of $f(Q,T)$ gravity models. This allows for the investigation of a wide range of wormhole configurations and properties within the context of $f(Q,T)$ gravity.\\
Now, let us take a moment to address the classical energy conditions derived from the Raychaudhuri equations. These conditions serve as a means to explore the physically plausible configurations of matter. Following is an expression for each of the four energy conditions: null energy condition (NEC), weak energy condition (WEC), dominant energy condition (DEC), and strong energy condition (SEC).\\
$\bullet$ Weak energy condition (\textbf{WEC}): $\rho\geq0$,\,\, $\rho+p_r\geq0$,\,\, and \,\, $\rho+p_t\geq0$.\\
$\bullet$ Null energy condition (\textbf{NEC}): $\rho+p_r\geq0$,\,\, and \,\, $\rho+p_t\geq0$.\\
$\bullet$ Dominant energy condition (\textbf{\textbf{DEC}}): $\rho\geq0$,\,\, $\rho+p_r\geq0$,\,\, $\rho+p_t\geq0$,\,\, $\rho-p_r\geq0$,\,\, and \,\, $\rho-p_t\geq0$.\\
$\bullet$ Strong energy condition (\textbf{SEC}): $\rho+p_r\geq0$,\,\, $\rho+p_t\geq0$,\,\, and \,\, $\rho+p_r+2p_t\geq0$.\\
To summarize, energy conditions serve as significant constraints on the behavior of matter within the Universe, playing a pivotal role in the investigation of wormholes. 

\section{Wormhole solutions with $f(Q, T)=\alpha\,Q+\beta\,T$}
\label{sec3}
In this section, we will examine a specific and notable $f(Q, T)$ model characterized by the equation:
\begin{equation}
\label{17}
f(Q, T)=\alpha\,Q+\beta\,T\,.
\end{equation}
Here, $\alpha$ and $\beta$ represent dimensionless model parameters. This model was introduced by Xu et al. \cite{Y.Xu} and naturally describes an exponentially expanding Universe. Loo et al. \cite{Loo1} employed the same model to investigate Bianchi type-I cosmology using observational datasets such as the Hubble parameter and Type Ia supernovae. 
Using the above linear functional form of 
$f(Q, T)$ i.e. Eq. \eqref{17} with constant redshift function, the field equations \eqref{14}-\eqref{16} can be reduced as
 \begin{equation}\label{18}
 \rho =\frac{\alpha b' (12 \pi -\beta )}{3 (\beta +8 \pi ) (4 \pi -\beta ) r^2}\,,
 \end{equation}
 \begin{equation}\label{19}
 p_r=-\frac{\alpha  \left(2 \beta  r b' +12 \pi  b -3 \beta  b\right)}{3 (\beta +8 \pi ) (4 \pi -\beta ) r^3}\,,
 \end{equation}
 \begin{equation}\label{20}
 p_t=-\frac{\alpha  \left((\beta +12 \pi ) r b'+3 b (\beta -4 \pi )\right)}{6 (4 \pi -\beta ) (\beta +8 \pi ) r^3}\,.
 \end{equation}
Now, we will study wormhole solutions under two DM halo profiles in the following subsections.
 \subsection{URC model}
 In this subsection, we will consider the URC model's energy density profile, which can be expressed using the following equation \cite{P. Salucci}
\begin{equation}\label{21}
\rho =\frac{\rho _{s} r_{s}^3}{(r+r_{s}) \left( r^2+r_{s}^2\right) },
\end{equation}
where $r_s$ and $\rho_s$ denote the characteristic radius and central density of URC dark matter halo, respectively.
Now, by comparing the Eqs. \eqref{18} and \eqref{21}, the differential equation under URC model is given by
\begin{equation}\label{22}
    \frac{\alpha  (12 \pi -\beta ) b'}{3 (4 \pi -\beta ) (\beta +8 \pi ) r^2}=\frac{\rho _{s} r_{s}^3}{(r+r_{s}) \left( r^2+r_{s}^2\right) }.
\end{equation}
On integrating the above eq. \eqref{22} for the shape function $b(r)$, we get
\begin{multline}\label{23}
b(r)= \frac{-\mathcal{K}_1}{48 \pi  \alpha -4 \alpha  \beta } \left(-\log \left(r^2+{r_s}^2\right)-2 \log (r+{r_s})
\right. \\ \left.
+2 \tan ^{-1}\left(\frac{r}{{r_s}}\right)\right)+c_1,
\end{multline}

where $c_1$ represents the integrating constant and $\mathcal{K}_1=3 (4 \pi -\beta ) (\beta +8 \pi ) \rho_s {r_s}^3$. Now, we impose the throat condition $b(r_0)=r_0$ to the above equation and obtain $c_1$ as
\begin{multline}\label{24}
  c_1=  \frac{\mathcal{K}_1 }{48 \pi  \alpha -4 \alpha  \beta }\left(-\log \left({r_0}^2+{r_s}^2\right)-2 \log ({r_0}+{r_s})
  \right. \\ \left.
  +2 \tan ^{-1}\left(\frac{{r_0}}{r_s}\right)\right)+r_0,
\end{multline}
where $r_0$ is the throat radius. Substituting the value of $c_1$ in Eq. \eqref{23}, we get,
\begin{multline}\label{25}
b(r)=r_0+ \frac{\mathcal{K}_1}{48 \pi  \alpha -4 \alpha  \beta } \left(-\log \left({r_0}^2+{r_s}^2\right)-2 \log ({r_0}
\right. \\ \left.
+{r_s})+2 \tan ^{-1}\left(\frac{{r_0}}{{r_s}}\right) + \log \left(r^2+{r_s}^2\right)+2 \log (r+{r_s})
\right. \\ \left.
-2 \tan ^{-1}\left(\frac{r}{{r_s}}\right)\right).
\end{multline}
The first plot of Fig. \ref{fig1} illustrates the graphical representation of the shape function along with all the essential properties required for the URC model. A detailed discussion about the behavior of the shape functions has been discussed in subsection- \ref{Diss}.\\
Now, substituting the shape function \eqref{25} in the Eqs. \eqref{19} and \eqref{20}, we could obtain the radial and tangential pressures given by 
\begin{multline}\label{25a}
p_r = \frac{\alpha}{\mathcal{M}_2}  \left(-3 (4 \pi -\beta ) \rho_s {r_s}^3 \mathcal{M}_1-\frac{8 \beta  \rho_s r^3 {r_s}^3}{(r+{r_s}) \left(r^2+{r_s}^2\right)}
\right. \\ \left.
+\frac{4 r_0 \alpha  (\beta -12 \pi )}{\beta +8 \pi }\right),
\end{multline}
\begin{multline}\label{25b}
p_t = \frac{\alpha}{2\mathcal{M}_2}\left(3 (4 \pi -\beta ) \rho_s {r_s}^3\mathcal{M}_1+4 r_0 \alpha  \left(\frac{20 \pi }{\beta +8 \pi }-1\right)
\right. \\ \left.
-\frac{4 (\beta +12 \pi ) \rho_s r^3 {r_s}^3}{(r+{r_s}) \left(r^2+{r_s}^2\right)}\right),
\end{multline}
where $\mathcal{M}_1=-\log \left(r_0^2+r_s^2\right)-2 \log (r_0+r_s)+2 \tan ^{-1}\left(\frac{r_0}{r_s}\right)+\log \left(r^2+{r_s}^2\right)+2 \log (r+r_s)-2 \tan ^{-1}\left(\frac{r}{r_s}\right)$ and $\mathcal{M}_2 = 2 r^3 (48 \pi  \alpha -4 \alpha  \beta )$.\\
Also, the NEC for radial and tangential pressures at the throat $r=r_0$ can be read as
\begin{multline}\label{25c}
\rho + p_r \bigg\vert_{r=r_0}= -\frac{3 (\beta -4 \pi ) \rho_s {r_s}^3}{(12 \pi -\beta ) (r_0+r_s) \left({r_0}^2+{r_s}^2\right)}\\
-\frac{\alpha }{{r_0}^2 (\beta +8 \pi )},
\end{multline}
\begin{multline}\label{25d}
\rho + p_t \bigg\vert_{r=r_0}=\frac{3 (4 \pi -\beta ) \rho_s {r_s}^3}{2(12 \pi -\beta ) (r_0+r_s) \left({r_0}^2+{r_s}^2\right)}\\
+\frac{\alpha }{{2\,r_0}^2 (\beta +8 \pi )}.
\end{multline}
It is evident that RHS of Eq. \eqref{25c} is a negative quantity which confirms the violation of NEC. The plots for radial and tangential NEC are shown in Figs. \ref{fig3} and \ref{fig4}. 
 \subsection{The cold dark matter (CDM) halo with NFW model-I and NFW model-II}
 The potential density presented by Hernquist \cite{L. Hernquist} was aimed at exploring the theoretical and observational aspects of elliptical galaxies. Subsequently, Navarro and his team \cite{J. F. Navarro} analyzed the equilibrium density profiles of DM halos in universes with hierarchical clustering using high-resolution N-body simulations. They found that the shape of these profiles remains consistent regardless of the halo mass, the spectral shape of the initial density fluctuations, or the values of the cosmological parameters. Two distinct CDM halo models for X-ray cluster halos and elliptical galaxies \cite{L. Hernquist, J. F. Navarro} are defined in \ref{subsubsection1} and \ref{subsubsection2}.
\subsubsection{NFW model-I}\label{subsubsection1}
First, we consider the energy density distribution for the NFW model-I, which can be defined as \cite{J. F. Navarro}
\begin{equation}\label{26}
\rho =\frac{\rho _{s} r_{s}}{r \left( \frac{r}{r_{s}}+1\right) ^2}\,,
\end{equation}
where $r_s$ and $\rho_s$ denote the characteristic radius and central density of the Universe, respectively.
Now, by comparing the Eqs. \eqref{18} and \eqref{26}, the differential equation under NFW model-I is given by
\begin{equation}\label{27}
\frac{\alpha  (12 \pi -\beta ) b'}{3 (4 \pi -\beta ) (\beta +8 \pi ) r^2}=\frac{\rho _{s} r_{s}}{r \left( \frac{r}{r_{s}}+1\right) ^2}.
\end{equation}
Integrating the preceding equation Eq. \eqref{27} for the shape function $b(r)$ yields
\begin{equation}\label{28}
b(r)= \frac{\mathcal{K}_1 \left((r+{r_s}) \log (r+{r_{s}})+{r_{s}}\right)}{\alpha  (12 \pi -\beta ) (r+{r_{s}})}+c_2,
\end{equation}

where $c_2$ is the integration constant. We next apply the throat condition to the aforementioned equation to get $c_2$ as
\begin{equation}\label{29}
  c_2=r_0-\frac{\mathcal{K}_1 \left((r_0+r_{s}) \log (r_0+r_{s})+r_{s}\right)}{\alpha  (12 \pi -\beta ) (r_0+r_{s})}.
\end{equation}
where $r_0$ is the throat radius. Thus, Eq. \eqref{28} can be modified by replacing the value of $c_2$, we get,
\begin{multline}\label{30}
b(r)=r_0+ \frac{\mathcal{K}_1 \left((r+{r_s}) \log (r+{r_{s}})+{r_{s}}\right)}{\alpha  (12 \pi -\beta ) (r+{r_{s}})}\\
-\frac{\mathcal{K}_1 \left((r_0+r_{s}) \log (r_0+r_{s})+r_{s}\right)}{\alpha  (12 \pi -\beta ) (r_0+r_{s})}.
\end{multline}
The behavior of the shape function is depicted graphically in the second plot of Fig. \ref{fig1}, along with all the essential properties needed for the NFW model-I.\\
The radial and tangential pressures are given by
\begin{equation}\label{30a}
p_r = -\frac{2 \beta  \rho_s {r_s}^3}{(12 \pi -\beta ) r (r+{r_s})^2}-\frac{r_0 \alpha }{r^3(\beta +8 \pi)}-\mathcal{M}_3 \,,
\end{equation}
\begin{equation}\label{30b}
    p_t = -\frac{(\beta +12 \pi ) \rho_s {r_s}^3}{2 (12 \pi -\beta ) r (r+{r_s})^2}+\frac{r_0 \alpha }{2 r^3(\beta +8 \pi)}+\frac{\mathcal{M}_3}{2} ,
\end{equation}
where $\mathcal{M}_3 = \frac{3 (4 \pi -\beta ) \rho_s {r_s}^3 (\log (r+{r_s})-\log (r_0+{r_s}))}{r^3 (12 \pi -\beta )}+\frac{3 (4 \pi -\beta ) \rho_s {r_s}^4 (r_0-r)}{r^3 (12 \pi -\beta ) (r_0+{r_s}) (r+{r_s})}$.\\
At wormhole throat, the radial and tangential NEC can be read as
\begin{equation}\label{1122}
    \rho + p_r \bigg\vert_{r=r_0} = -\frac{\alpha }{{r_0}^2 (\beta +8 \pi) }-\frac{2  \beta  \rho_s {r_s}^3}{r_0 (12 \pi -\beta ) (r_0+{r_s})^2} ,
\end{equation}
\begin{equation}\label{1133}
    \rho + p_t \bigg\vert_{r=r_0} = \frac{\alpha }{2 {r_0}^2 (\beta +8 \pi )}-\frac{ (\beta +12 \pi ) \rho_s {r_s}^3}{2r_0 (12 \pi -\beta ) (r_0+{r_s})^2}.
\end{equation}
It is confirmed from the above expressions that the Eq. \eqref{1122} is negative, whereas the Eq. \eqref{1133} is a positive quantity. Therefore, mathematically, we can conclude that NEC is disrespected in the neighborhood of the throat. Moreover, we have presented the graphical view in Figs. \ref{fig3} and \ref{fig4}.
\subsubsection{(NFW) model-II}\label{subsubsection2}
Again, the other NFW model's energy density profile can be expressed using the following equation
\begin{equation}\label{31}
\rho =\frac{\rho _{s} r_{s}}{r \left( \frac{r}{r_{s}}+1\right) ^3},
\end{equation}

\begin{figure*}[t]
\centering
\includegraphics[width=17cm,height=5cm]{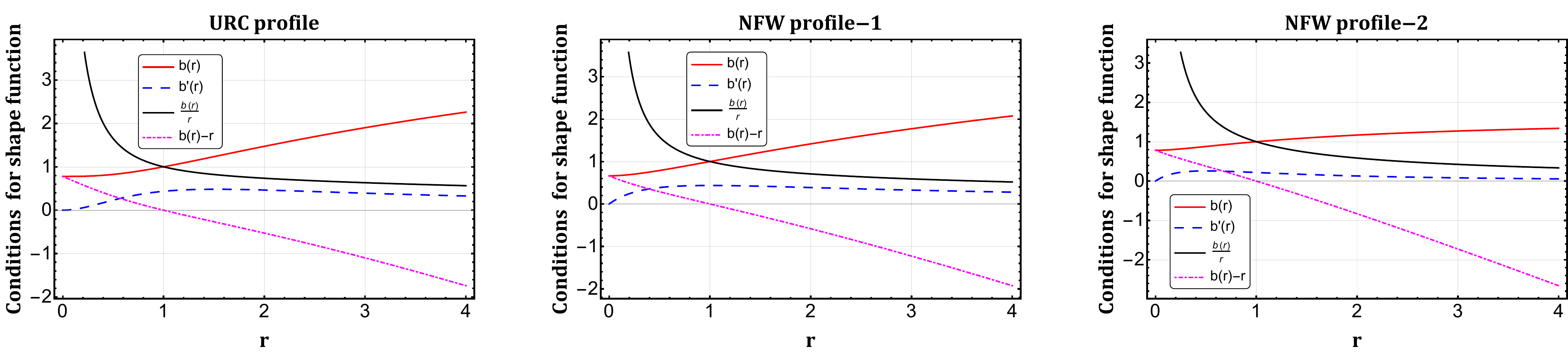}
\caption{profile shows the conditions of shape function with respect to $r$. We consider $\alpha=1.2,\,\beta=5,\,\rho_0=0.1,\, r_s=1\, \text{and} \, r_0 = 1$.}
\label{fig1}
\end{figure*}

Now, by comparing the Eqs. \eqref{18} and \eqref{31}, the differential equation under NFW model-II is given by
\begin{equation}\label{32}
    \frac{\alpha  (12 \pi -\beta ) b'}{3 (4 \pi -\beta ) (\beta +8 \pi ) r^2}=\frac{\rho _{s} r_{s}}{r \left( \frac{r}{r_{s}}+1\right) ^3}\,.
\end{equation}
By integrating the above differential Eq. \eqref{32} for the shape function $b(r)$, we get
\begin{equation}\label{33}
   b(r)=-\frac{\mathcal{K}_1 r_s (2 r+r_s)}{2 \alpha  (12 \pi -\beta ) (r+r_s)^2}+c_3\,,
\end{equation}

where $c_3$ represents the constant of integration. Now we use the throat condition to the above equation and obtain $c_3$ as
\begin{equation}\label{34}
  c_3=r_0+\frac{\mathcal{K}_1 r_s (2 r_0+r_s)}{2 \alpha  (12 \pi -\beta ) (r_0+r_s)^2}\,,
\end{equation}
where $r_0$ is the throat radius. Substituting the value of $c_3$ in Eq. \eqref{33}, we get,
\begin{equation}\label{35}
    b(r)=r_0+\frac{\mathcal{K}_1 r_s (2 r_0+r_s)}{2 \alpha  (12 \pi -\beta ) (r_0+r_s)^2}
    -\frac{\mathcal{K}_1 r_s (2 r+r_s)}{2 \alpha  (12 \pi -\beta ) (r+r_s)^2}\,.
\end{equation}
The above shape function satisfies the asymptotic flatness condition with some appropriate fixed parameters. Also, we checked one of the essential properties of shape function, i.e., flare-out condition. We obtained the specific range of model parameter $\beta$, where flare-out is satisfied under asymptotic background. In the third plot of Fig. \ref{fig1}, we have presented the behaviors of the shape function and their detailed discussions are written in section- \ref{Diss}.\\
For this case, the NEC for radial and tangential pressures can be read as
\begin{multline}\label{35a}
\rho + p_r = -\frac{2 r_0 \alpha }{2 r^3 (\beta +8 \pi )}+\frac{1}{\mathcal{M}_4} \left( 3 (4 \pi -\beta ) \rho_s {r_s}^4 \left({r_s}^2
\right.\right. \\ \left.\left.
\left({r_0}^2+r^2\right)+2 r_0 r^2 (2 r_0-r)+r {r_s} (3 r_0-r) (r_0+r)\right) \right)\,,
\end{multline}
\begin{multline}\label{35b}
\rho + p_t = \frac{2 r_0 \alpha }{4r^3 (\beta +8 \pi) } - \frac{1}{2 \mathcal{M}_4} \left( 3 (4 \pi -\beta ) \rho_s {r_s}^4 \left({r_0}^2 {r_s} 
\right.\right. \\ \left.\left.
(3 r+{r_s})-2 r_0 r^2 (r+3 {r_s})-r^2 {r_s} (r+3 {r_s})\right)\right)\,,
\end{multline}
where $\mathcal{M}_4 = 2 r^3 (12 \pi -\beta ) (r_0+r_s)^2 (r+r_s)^3$.\\
Now, at the throat $r=r_0$, the above expressions for NEC will reduce to
\begin{equation}\label{35c}
\rho + p_r \bigg\vert_{r=r_0} = \frac{3 (4 \pi -\beta ) \rho_s {r_s}^4}{{r_0} (12 \pi -\beta ) ({r_0}+{r_s})^3}-\frac{\alpha }{{r_0}^2 (\beta +8 \pi) } \,,
\end{equation}
\begin{equation}\label{35d}
\rho + p_t \bigg\vert_{r=r_0} = \frac{\alpha }{2 {r_0}^2 (\beta +8 \pi )}+\frac{3 (4 \pi -\beta ) \rho_s {r_s}^4}{2 {r_0} (12 \pi -\beta ) ({r_0}+{r_s})^3}  \,.
\end{equation}
The graphical evolution of NEC for this NFW model-II profile is depicted in Figs. \ref{fig3} and \ref{fig4}. Moreover, the remaining energy conditions are also discussed in Section-\ref{Diss}.
\subsection{Discussions}
\label{Diss}
\begin{figure*}[t]
\centering
\includegraphics[width=17cm,height=5cm]{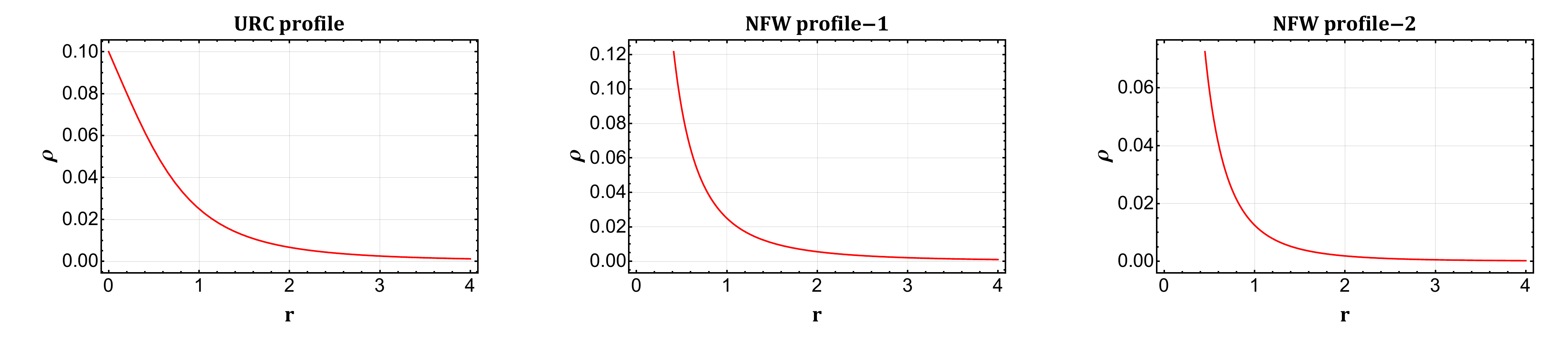}
\caption{Profile shows the behavior of energy density $\rho$ with respect to $r$. We consider $\rho_0=0.1,\,  \text{and} \, r_s=1$.}
\label{fig2}
\end{figure*}

\begin{figure*}[t]
\centering
\includegraphics[width=17cm,height=5cm]{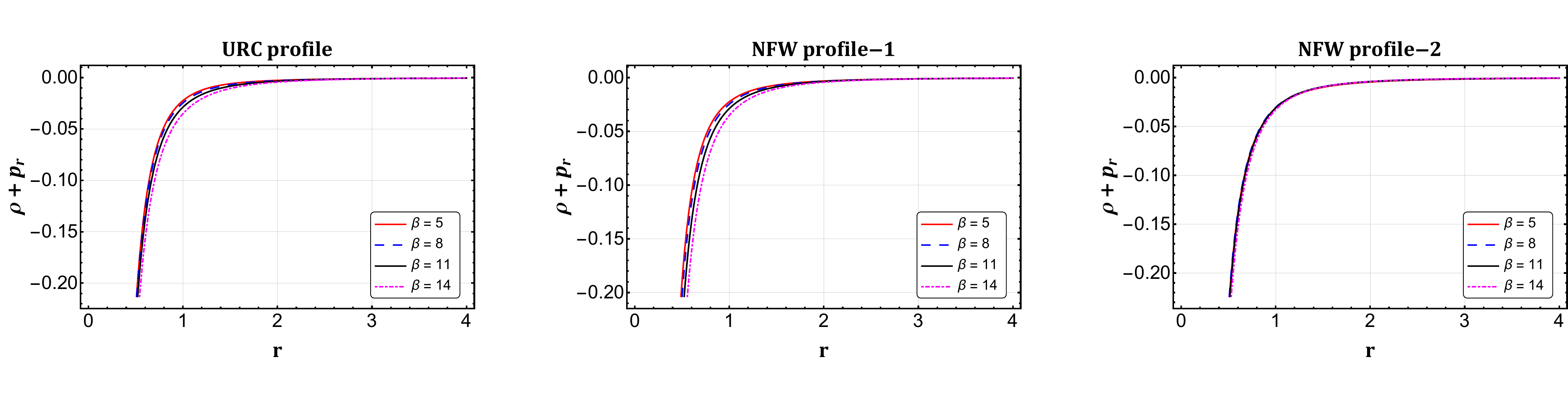}
\caption{Profile shows the behavior of NEC for radial pressure $\rho+p_r$ with respect to $r$ for different values of $\beta$. We consider $\alpha=1.2,\, \rho_0=0.1,\, r_s=1,\, \text{and} \, r_0 = 1$.}
\label{fig3}
\end{figure*}

\begin{figure*}[t]
\centering
\includegraphics[width=17cm,height=5cm]{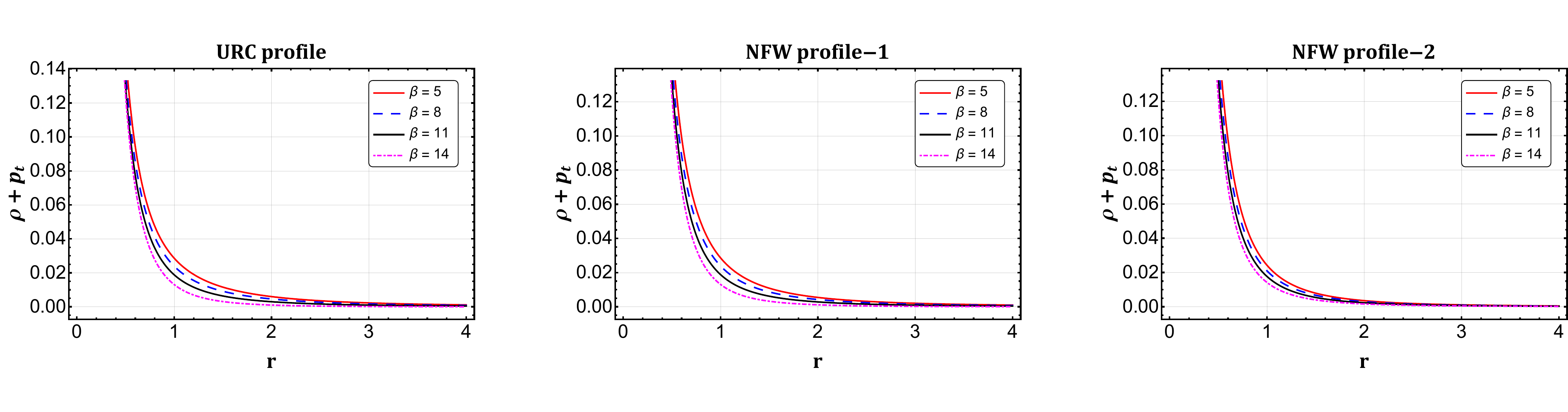}
\caption{Profile shows the behavior of NEC for tangential pressure $\rho+p_t$ with respect to $r$ for different values of $\beta$. We consider $\alpha=1.2,\, \rho_0=0.1,\, r_s=1,\, \text{and} \, r_0 = 1$.}
\label{fig4}
\end{figure*}

\begin{figure*}[t]
\centering
\includegraphics[width=17cm,height=5cm]{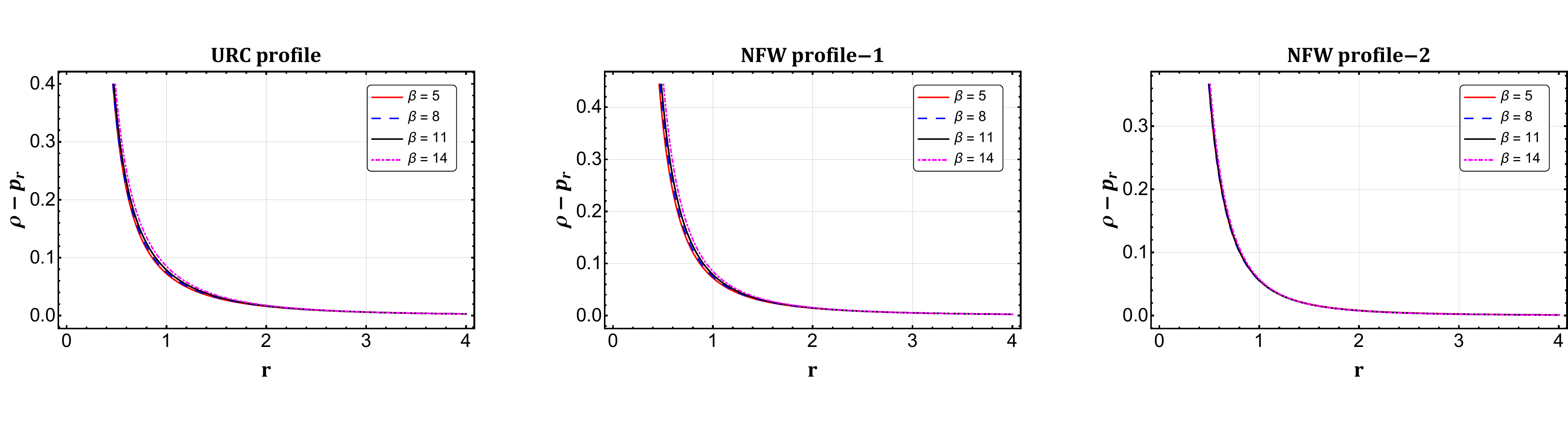}
\caption{Profile shows the behavior of DEC for radial pressure $\rho-p_r$ with respect to $r$ for different values of $\beta$. We consider $\alpha=1.2,\, \rho_0=0.1,\, r_s=1,\, \text{and} \, r_0 = 1$.}
\label{fig5}
\end{figure*}

\begin{figure*}[t]
\centering
\includegraphics[width=17cm,height=5cm]{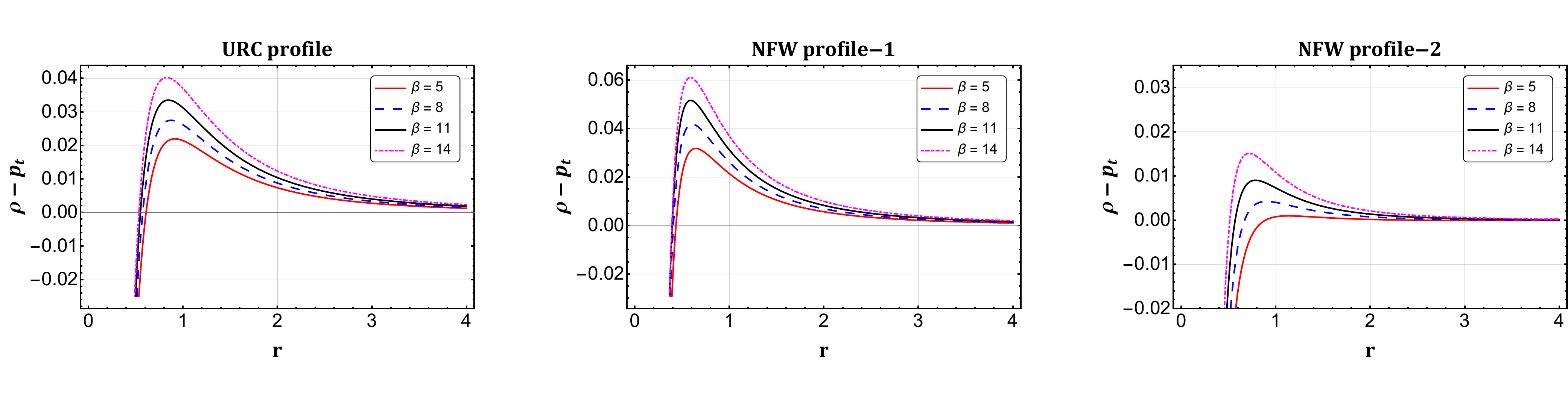}
\caption{Profile shows the behavior of DEC for tangential pressure $\rho-p_t$ with respect to $r$ for different values of $\beta$. We consider $\alpha=1.2,\, \rho_0=0.1,\, r_s=1,\, \text{and} \, r_0 = 1$.}
\label{fig6}
\end{figure*}

\begin{figure*}[t]
\centering
\includegraphics[width=17cm,height=5cm]{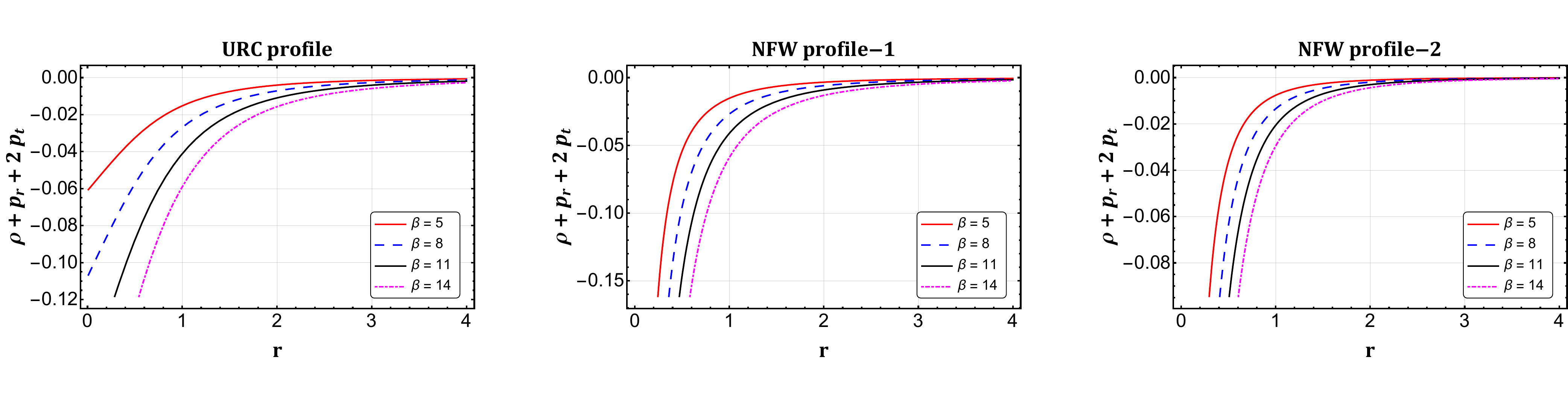}
\caption{Profile shows the behavior of SEC $\rho+p_r+2p_t$ with respect to $r$ for different values of $\beta$. We consider $\alpha=1.2,\, \rho_0=0.1,\, r_s=1,\, \text{and} \, r_0 = 1$.}
\label{fig7}
\end{figure*}

In this particular subsection, we will discuss the behavior of obtained shape functions as well as energy conditions for both URC and NFW models. Let us first discuss the obtained shape functions, which can be found in Eqs. \eqref{25}, \eqref{30}, and \eqref{35} for URC, NFW model-I and NFW model-II, respectively. From the expressions of the shape function, it is clear that $\alpha\neq 0$ and $\beta\neq 12\pi$. Also, we fixed some free parameters such as $\alpha=1.2,\,\rho_0=0.1,\, r_s=1\, \text{and} \, r_0 = 1$ depending on the study of shape functions. We checked the flare-out condition at the wormhole throat and noticed that at $r=r_0$, the flare-out condition $b^{'}(r_0)<1$ is satisfied. In that case, the model parameter $\beta$ should be less than $37.6991$, i.e., $\beta < 37.6991$ for each case. Considering the range of $\beta$ into account, we have studied the asymptotic flatness conditions for both URC and NFW models. It indicates that as the radial distance increases, the ratio $\frac{b(r)}{r}$ approaches zero, ensuring the satisfaction of the asymptotic behavior of the shape function. The graphical representation of all the necessary properties of the shape functions for each model has been shown in Fig. \ref{fig1}.\\
Later, we discussed energy conditions for both URC and NFW models with the appropriate choices of free parameters. We have depicted the plot for energy density versus radial coordinate $r$ in Fig. \ref{fig2}, which shows positively decreasing behavior in the entire space-time. Also, in Figs. \ref{fig3}, we have shown the behavior of $\rho+p_r$, indicates the violation of radial NEC. In this case, the range of $\beta$ should be $-25.1327<\beta <19.2283$ (URC model), $-25.1327<\beta <19.2283$ (NFW model-I) and $-25.1327<\beta <26.6957$ (NFW model-II). Moreover, within this particular range, we checked the behavior of $\rho+p_t$ and found that tangential NEC is satisfied (see Fig. \ref{fig4}.) Further, we have investigated the DEC and SEC for different values of $\beta$ in Figs. \ref{fig5},  \ref{fig6}, and \ref{fig7}, respectively. One can check that DEC is satisfied, whereas SEC is violated in the vicinity of the throat. Furthermore, one can see Table- \ref{Table 1} for the overview of the energy conditions at the throat.


\begin{table*}[t]
\begin{center}
\begin{tabular}{|c|c|c|c|c|}
\hline
Energy conditions  & URC profile & NFW profile-1 & NFW profile-2  \\ \hline
$\rho$  & $satisfied$ & $satisfied$ & $satisfied$   \\ 
\hline
$\rho + p_r$ & $violated$ & $violated$ & $violated$   \\ 
\hline
$\rho + p_t$ & $satisfied$ & $satisfied$ & $satisfied$   \\ 
\hline
$\rho - p_r$ & $satisfied$ & $satisfied$ & $satisfied$   \\ 
\hline
$\rho - p_t$ & $satisfied$ & $satisfied$ & $satisfied$   \\ 
\hline
$\rho + p_r + 2 p_t$ & $violated$ & $violated$ & $violated$   \\ 
\hline
\end{tabular}
 \caption{Description of Linear Model}
\label{Table 1}
\end{center}
\end{table*}

\section{Wormhole solutions with $f(Q, T)=\alpha_1 +\beta_1  \log (Q)+\gamma  T$}
\label{sec4}
In this section, we shall consider a specific and interesting $f(Q,T)$ model, which is the extension of $f(Q)$ model \cite{Sanjay 1} given by 
\begin{equation}
\label{36}
f(Q, T)=\alpha_1 +\beta_1  \log (Q)+\gamma  T\,,
\end{equation}
 where $\alpha_1$, $\beta_1$, and $\gamma$ are model parameters. Using the above non-linear functional form \eqref{36}, the field equations are obtained as

 \begin{multline}\label{37}
\rho = \frac{1}{\mathcal{K}_2}\left[-10 \beta  \gamma  b' \phi ' \left(b'+2 r \phi '\right) r^5+5 \beta  \gamma  b \left(\phi ' \left(b'^2+2 
\right.\right.\right. \\ \left.\left.\left.
\left(6 r \phi '+7\right) b'+4 r \phi ' \left(r \phi '+4\right)-2 r b''\right)+2 r b' \phi ''\right) r^4+b^2 
\right. \\ \left.
\left(6 \beta  (16 \pi -3 \gamma ) \phi '' r^2+4 \left(\phi ' \left(3 r \phi ' \left(\gamma  \alpha +8 \pi  \alpha -8 \pi  \beta -21 \beta  
\right.\right.\right.\right.\right. \\ \left.\left.\left.\left.\left.
\gamma +\beta  (\gamma +8 \pi ) \log \left(\mathcal{K}_3\right)-5 r \beta  \gamma  \phi '\right)-\beta  (11 \gamma +48 \pi )\right)+\beta 
\right.\right.\right. \\ \left.\left.\left.
  \left(5 r \phi ' \gamma -\gamma +12 \pi \right) b''\right) r+b'^2 \left(24 \pi  (\alpha +\beta )+(3 \alpha -7 \beta ) \gamma 
\right.\right.\right. \\ \left.\left.\left.
 +3 \beta (\gamma +8 \pi ) \log \left(\mathcal{K}_3\right)+5 r \beta  \gamma  \phi '\right)+2 b' \left(8 \beta  (\gamma -12 \pi )+r 
 \right.\right.\right. \\ \left.\left.\left.
 \left(\phi ' \left(6 \gamma  \alpha +48 \pi  \alpha -48 \pi  \beta -61 \beta  \gamma +6 \beta  (\gamma +8 \pi ) \log \left(\mathcal{K}_3\right)
 \right.\right.\right.\right.\right. \\ \left.\left.\left.\left.\left.
 -30 r \beta  \gamma  \phi '\right)-10 r \beta  \gamma  \phi ''\right)\right)\right) r^2+b^3 \left(60 r^3 \beta  \gamma  \phi '^3-24 r^2 
  \right.\right. \\ \left.\left.
 \left(\gamma  \alpha +8 \pi  \alpha -8 \pi  \beta -11 \beta  \gamma +\beta  (\gamma +8 \pi ) \log \left(\mathcal{K}_3\right)\right) \phi '^2-r 
  \right.\right. \\ \left.\left.
 \left(96 \pi  (\alpha -5 \beta )+(12 \alpha -85 \beta ) \gamma +12 \beta  (\gamma +8 \pi ) \log \left(\mathcal{K}_3\right)+
  \right.\right.\right. \\ \left.\left.\left.
 10 r \beta  \gamma  b''\right) \phi ')-4 \beta  (12 \pi -\gamma ) \left(r b''-4\right)+12 r^2 \beta  (3 \gamma -16 \pi )
  \right.\right. \\ \left.\left.
 -16 \pi ) \phi ''-2 b' \left(24 \pi  (\alpha -3 \beta )+(3 \alpha +\beta ) \gamma +3 \beta  (\gamma 
  \right.\right.\right. \\ \left.\left.\left.
 \log \left(\mathcal{K}_3\right)+r \left(2 \phi ' \left(3 \gamma  \alpha +24 \pi  \alpha -24 \pi  \beta -13 \beta  \gamma +3 \beta 
 \right.\right.\right.\right.\right. \\ \left.\left.\left.\left.\left.
 (\gamma +8 \pi ) \log \left(\mathcal{K}_3\right)-5 r \beta  \gamma  \phi '\right)-5 r \beta  \gamma  \phi ''\right)\right)\right) r+b^4 
  \right. \\ \left.
 \left(3 \left(8 \pi  (\alpha -7 \beta )+(\alpha +3 \beta ) \gamma +\beta  (\gamma +8 \pi ) \log \left(\mathcal{K}_3\right)\right)+
  \right.\right. \\ \left.\left.
 r \left(\phi ' \left(96 \pi  (\alpha -3 \beta )+(12 \alpha -41 \beta ) \gamma +12 \beta  (\gamma +8 \pi ) 
 \right.\right.\right.\right. \\ \left.\left.\left.\left.
 \log \left(\mathcal{K}_3\right)+4 r \phi ' \left(3 \gamma  \alpha +24 \pi  \alpha -24 \pi  \beta -23 \beta  \gamma +3 \beta  (\gamma 
 \right.\right.\right.\right.\right. \\ \left.\left.\left.\left.\left.
 +8 \pi ) \log \left(\mathcal{K}_3\right)-5 r \beta  \gamma  \phi '\right)\right)+6 r \beta  (16 \pi -3 \gamma ) \phi ''\right)\right)\right]\,,
 \end{multline}
 \begin{multline}\label{38}
p_r = \frac{-1}{\mathcal{K}_2}\left[-10 \beta  \gamma  b' \phi ' \left(b'+2 r \phi '\right) r^5+\beta  b \left(2 \phi ' \left(2 \phi ' 
 \right.\right.\right. \\ \left.\left.\left.
\left(8 (\gamma +6 \pi )+5 r \gamma  \phi '\right)-5 \gamma  b''\right) r^2+2 b' \left(\phi ' \left(30 r \phi ' \gamma +
\right.\right.\right.\right. \\ \left.\left.\left.\left.
11 \gamma +96 \pi \right)+5 r \gamma  \phi ''\right) r+b'^2 \left(5 r \phi ' \gamma -12 \gamma +48 \pi \right)\right) r^3
\right. \\ \left.
+b^2 \left(6 \beta  (16 \pi -3 \gamma ) \phi '' r^2+4 \left(\phi ' \left(\beta  (\gamma -96 \pi )+3 r \phi ' 
\right.\right.\right.\right. \\ \left.\left.\left.\left.
\left(8 \pi  (\alpha -7 \beta )+(\alpha -9 \beta ) \gamma +\beta  (\gamma +8 \pi ) \log \left(\mathcal{K}_3\right)-5 r \beta  \gamma 
\right.\right.\right.\right.\right. \\ \left.\left.\left.\left.\left.
\phi '\right)\right)+\beta  \left(5 r \phi ' \gamma -\gamma +12 \pi \right) b''\right) r+b'^2 \left(3 \gamma  \alpha +24 \pi  \alpha
 \right.\right.\right. \\ \left.\left.\left.
-24 \pi  \beta +5 \beta  \gamma +3 \beta  (\gamma +8 \pi ) \log \left(\mathcal{K}_3\right)+5 r \beta  \gamma  \phi '\right)+2 b' 
\right.\right. \\ \left.\left.
\left(4 \beta  (5 \gamma -36 \pi )+r \left(\phi ' \left(48 \pi  (\alpha -5 \beta )+(6 \alpha -13 \beta ) \gamma +
\right.\right.\right.\right.\right. \\ \left.\left.\left.\left.\left.
6 \beta  (\gamma +8 \pi ) \log \left(\mathcal{K}_3\right)-30 r \beta  \gamma  \phi '\right)-10 r \beta  \gamma  \phi ''\right)\right)\right) r^2
\right. \\ \left.
-b^3 \left(-60 r^3 \beta  \gamma  \phi '^3+24 r^2 \left(8 \pi  (\alpha -4 \beta )+(\alpha -5 \beta ) \gamma +\beta 
 \right.\right.\right. \\ \left.\left.\left.
(\gamma +8 \pi ) \log \left(\mathcal{K}_3\right)\right) \phi '^2+r \left(96 \pi  (\alpha -9 \beta )+(12 \alpha +11 \beta )
 \right.\right.\right. \\ \left.\left.\left.
\gamma +12 \beta  (\gamma +8 \pi ) \log \left(\mathcal{K}_3\right)+10 r \beta  \gamma  b''\right) \phi '+4 \beta  \left(7 \gamma +r 
 \right.\right.\right. \\ \left.\left.\left.
\left((12 \pi -\gamma ) b''+3 r (16 \pi -3 \gamma ) \phi ''\right)-60 \pi \right)+2 b' \left(24 \pi
 \right.\right.\right. \\ \left.\left.\left.
(\alpha -5 \beta )+(3 \alpha +13 \beta ) \gamma +3 \beta  (\gamma +8 \pi ) \log \left(\mathcal{K}_3\right)+r \left(2 
\right.\right.\right.\right. \\ \left.\left.\left.\left.
\phi ' \left(3 \gamma  \alpha +24 \pi  \alpha -72 \pi  \beta -\beta  \gamma +3 \beta  (\gamma +8 \pi ) \log \left(\mathcal{K}_3\right)
\right.\right.\right.\right.\right. \\ \left.\left.\left.\left.\left.
-5 r \beta  \gamma  \phi '\right)-5 r \beta  \gamma  \phi ''\right)\right)\right) r+b^4 \left(3 \left(8 \pi  (\alpha -9 \beta )+(\alpha 
 \right.\right.\right. \\ \left.\left.\left.
+7 \beta ) \gamma +\beta  (\gamma +8 \pi ) \log \left(\mathcal{K}_3\right)\right)+r \left(\phi ' \left(96 \pi  (\alpha -5 \beta )+
\right.\right.\right.\right. \\ \left.\left.\left.\left.
(12 \alpha +7 \beta ) \gamma +12 \beta  (\gamma +8 \pi ) \log \left(\mathcal{K}_3\right)+4 r \phi ' \left(3 \gamma  \alpha +24 
\right.\right.\right.\right.\right. \\ \left.\left.\left.\left.\left.
\pi  \alpha -72 \pi  \beta -11 \beta  \gamma +3 \beta  (\gamma +8 \pi ) \log \left(\mathcal{K}_3\right)-5 r \beta  \gamma  \phi '\right)\right)
 \right.\right.\right. \\ \left.\left.\left.
+6 r \beta  (16 \pi -3 \gamma ) \phi ''\right)\right)\right]\,,
 \end{multline}
 \begin{multline}\label{39}
p_t = \frac{1}{\mathcal{K}_2}\left[2 \beta  (24 \pi -\gamma ) b' \phi ' \left(b'+2 r \phi '\right) r^5-\beta  b \left(2 \phi ' 
\right.\right. \\ \left.\left.
\left(2 \phi ' \left(2 (\gamma +36 \pi )+r (24 \pi -\gamma ) \phi '\right)+(\gamma -24 \pi ) b''\right) r^2
\right.\right. \\ \left.\left.
+2 b' \left(\phi ' \left(5 (\gamma +24 \pi )+6 r (24 \pi -\gamma ) \phi '\right)+r (24 \pi -\gamma )
 \right.\right.\right. \\ \left.\left.\left.
\phi ''\right) r+b'^2 \left(6 (\gamma -4 \pi )+r (24 \pi -\gamma ) \phi '\right)\right) r^3+b^2 \left(6 \beta 
\right.\right. \\ \left.\left.
(8 \pi -3 \gamma ) \phi '' r^2+4 \left(\phi ' \left(\beta  (11 \gamma +48 \pi )+3 r \phi ' \left(-\gamma  \alpha -
\right.\right.\right.\right.\right. \\ \left.\left.\left.\left.\left.
8 \pi  \alpha +80 \pi  \beta +3 \beta  \gamma -\beta  (\gamma +8 \pi ) \log \left(\mathcal{K}_3\right)+r \beta  (24 \pi -
\right.\right.\right.\right.\right. \\ \left.\left.\left.\left.\left.
\gamma ) \phi '\right)\right)+\beta  \left(r (\gamma -24 \pi ) \phi '-2 \gamma \right) b''\right) r+b'^2 \left(-3 \gamma  \alpha
 \right.\right.\right. \\ \left.\left.\left.
-24 \pi  \alpha +\beta  \gamma -3 \beta  (\gamma +8 \pi ) \log \left(\mathcal{K}_3\right)+r \beta  (\gamma -24 \pi ) \phi '\right)
\right.\right. \\ \left.\left.
+2 b' \left(22 \gamma  \beta -24 \pi  \beta +r \left(\phi ' \left(-6 \gamma  \alpha -48 \pi  \alpha +216 \pi  \beta +
\right.\right.\right.\right.\right. \\ \left.\left.\left.\left.\left.
19 \beta  \gamma -6 \beta  (\gamma +8 \pi ) \log\left(\mathcal{K}_3\right)+6 r \beta  (24 \pi -\gamma ) \phi '\right)+2 r \beta  
\right.\right.\right.\right. \\ \left.\left.\left.\left.
(24 \pi -\gamma ) \phi ''\right)\right)\right) r^2+b^3 \left(-38 \gamma  \beta +24 \pi  \beta +r \left(12 r^2 \beta 
 \right.\right.\right. \\ \left.\left.\left.
(\gamma -24 \pi ) \phi '^3+24 r \left(\gamma  \alpha +8 \pi  \alpha -44 \pi  \beta -2 \beta  \gamma +\beta  (\gamma +
\right.\right.\right.\right. \\ \left.\left.\left.\left.
8 \pi ) \log \left(\mathcal{K}_3\right)\right) \phi '^2+\left(12 \gamma  \alpha +96 \pi  \alpha -408 \pi  \beta -103 \beta  \gamma +
\right.\right.\right.\right. \\ \left.\left.\left.\left.
12 \beta  (\gamma +8 \pi ) \log \left(\mathcal{K}_3\right)+2 r \beta  (24 \pi -\gamma ) b''\right) \phi '+8 \beta  \gamma  b''
 \right.\right.\right. \\ \left.\left.\left.
+12 r \beta  (3 \gamma -8 \pi ) \phi ''\right)+2 b' \left(3 \gamma  \alpha +24 \pi  \alpha -17 \beta  \gamma +3 \beta 
 \right.\right.\right. \\ \left.\left.\left.
(\gamma +8 \pi ) \log \left(\mathcal{K}_3\right)+r \left(2 \phi ' \left(3 \gamma  \alpha +24 \pi  \alpha -48 \pi  \beta -7 \beta  \gamma 
\right.\right.\right.\right.\right. \\ \left.\left.\left.\left.\left.
+3 \beta  (\gamma +8 \pi ) \log \left(\mathcal{K}_3\right)+r \beta  (\gamma -24 \pi ) \phi '\right)+r \beta  (\gamma -24 
\right.\right.\right.\right. \\ \left.\left.\left.\left.
\pi ) \phi ''\right)\right)\right) r-b^4 \left(3 \left(\gamma  \alpha +8 \pi  \alpha -11 \beta  \gamma +\beta  (\gamma +8 \pi ) 
 \right.\right.\right. \\ \left.\left.\left.
\log \left(\mathcal{K}_3\right)\right)+r \left(\phi ' \left(12 \gamma  \alpha +96 \pi  \alpha -216 \pi  \beta -59 \beta  \gamma +12 
\right.\right.\right.\right. \\ \left.\left.\left.\left.
\beta  (\gamma +8 \pi ) \log \left(\mathcal{K}_3\right)+4 r \phi ' \left(3 \gamma  \alpha +24 \pi  \alpha -96 \pi  \beta -5 \beta  \gamma 
\right.\right.\right.\right.\right. \\ \left.\left.\left.\left.\left.
+3 \beta  (\gamma +8 \pi ) \log \left(\mathcal{K}_3\right)+r \beta  (\gamma -24 \pi ) \phi '\right)\right)+6 r \beta  (3 \gamma 
 \right.\right.\right. \\ \left.\left.\left.
-8 \pi ) \phi ''\right)\right)\right]\,,
 \end{multline}
 where $\mathcal{K}_2 = 12 (4 \pi -\gamma ) (\gamma +8 \pi ) b^2 \left(b-r b'+2 r (b-r) \phi '\right)^2$ and $\mathcal{K}_3 = \frac{b \left(b-r b'+2 r (b-r) \phi '\right)}{r^3 (r-b)}$.
 
In the current investigation, we will use the Karmarkar condition \cite{Karmarkar/1948} with embedded class-1 space-time to find wormhole solutions. One of the most important aspects of the current analysis is this condition. The embedded class-1 solution of Riemannian space is necessary for the basic formulation of the Karmarkar condition. Eisenhart provided an essential and appropriate requirement for the embedded class-1 solution \cite{Eisenhart/1966}, which depends on the Riemann curvature tensor, $\mathcal{R}_{mnpq}$, and on a symmetric tensor of the second order, $b_{mn}$, through

\begin{itemize}
\item the Gauss equation:
\begin{eqnarray}\label{eqcls1.1}
\mathcal{R}_{mnpq}=2\,\epsilon\,{b_{m\,[p}}{b_{q]n}}\,,
\end{eqnarray}
\item the Codazzi equation:
\begin{eqnarray}\label{eqcls1.2}
b_{m\left[n;p\right]}-{\Gamma}^q_{\left[n\,p\right]}\,b_{mq}+{{\Gamma}^q_{m}}\,{}_{[n}\,b_{p]q}=0.
\end{eqnarray}
\end{itemize}
Here, we consider the case where $\epsilon=\pm1$, and square brackets denote antisymmetrization. The coefficients of the second differential form are represented by $b_{mn}$. By utilizing Eqs.~(\ref{eqcls1.1}) and ~(\ref{eqcls1.2}) and applying the prescribed mathematical procedure, we can calculate the Karmarkar condition as follows:
\begin{equation}\label{40}
\mathcal{R}_{2323}\mathcal{R}_{1414}=\mathcal{R}_{1224}\mathcal{R}_{1334}+ \mathcal{R}_{1212}\mathcal{R}_{3434},
\end{equation}
with Pandey and Sharma condition \cite{Pandey/1981}, i.e., $\mathcal{R}_{2323}\neq\mathcal{R}_{1414}\neq0$.

Through the substitution of the suitable Riemannian tensor components into equation (\ref{40}), the ensuing result yields the following differential equation:
\begin{multline}\label{41}
\hspace{-0.8cm} \frac{\nu'(r) \lambda'(r)}{1-e^{\lambda(r)}}-\left\{\nu'(r) \lambda'(r)+\nu'(r)^2-2 \left[\nu''(r)+\nu'(r)^2\right]\right\}=0,\\
e^{\lambda(r)}\neq1,
\end{multline}
Due to the embedded class-I solution, we assume the following redshift function \cite{Zia,Anchordoqui}
\begin{equation}\label{44}
\nu(r)=2\phi(r)=-\frac{2\chi }{r}, \,\,\, \chi>0.
\end{equation}
The above redshift function satisfies the flatness condition, i.e., $\nu(r) \rightarrow 0$ when $r \rightarrow \infty$.\\
On solving Eq. (\ref{41}), we obtain
\begin{equation}\label{42}
e^{\lambda(r)}=1+Ae^{\nu(r)}\nu^{'2}(r),
\end{equation}
where $A$ is the constant of integration. Now, following the procedure written in \cite{Mustafa11}, produces the below embedded shape function
\begin{equation}\label{46}
b(r)=r -\frac{\delta  r^5}{r_0^4 (r_0-\delta ) e^{-2 \chi  \left(\frac{1}{r}-\frac{1}{r_0}\right)}+\delta  r^4}+\delta ,\;\; 0<\delta<r_{0}.
\end{equation}

\begin{figure}[h]
\centering
\includegraphics[width=5cm,height=5cm]{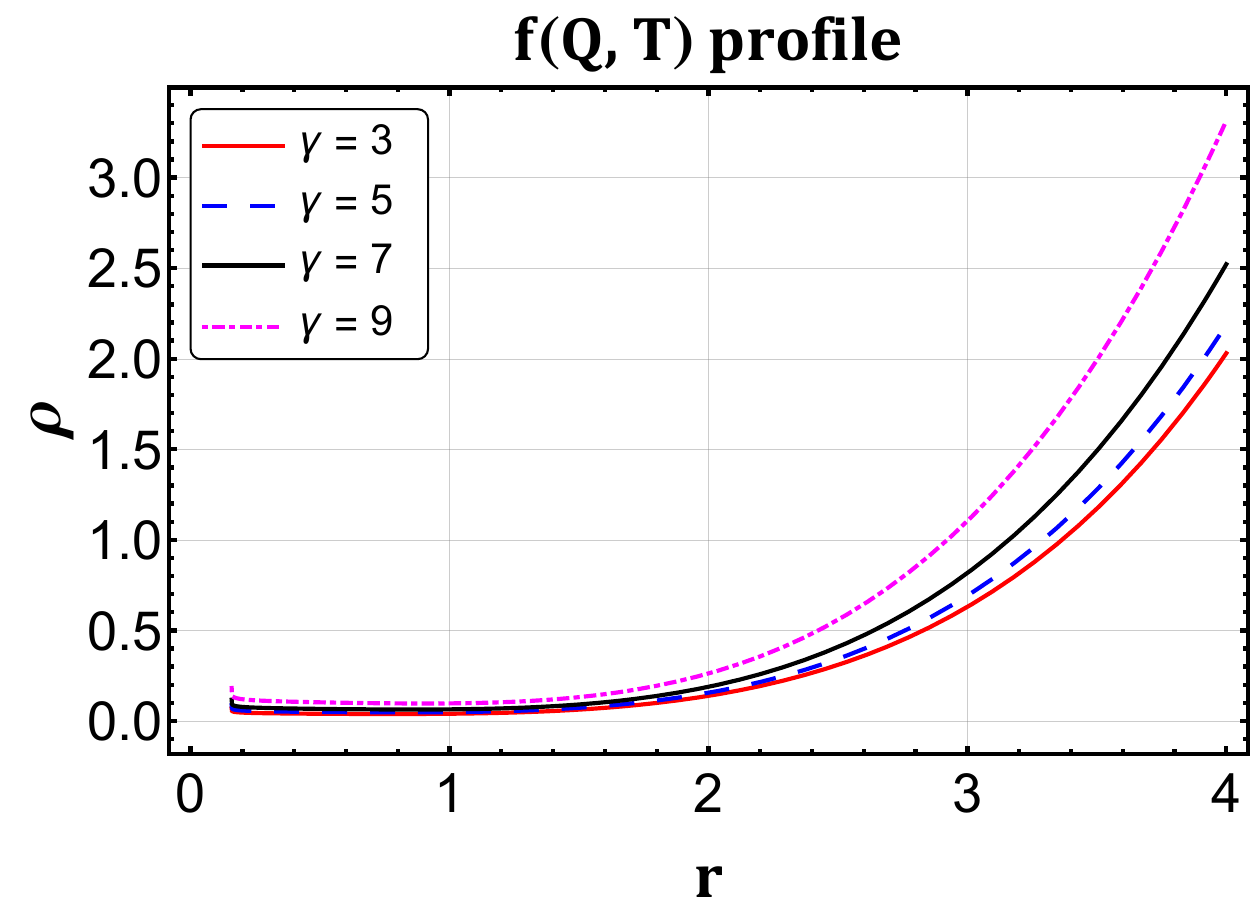}
\caption{Profile shows the behavior of energy density $\rho$ with respect to $r$ for different values of $\gamma$. We consider $r_0=1,\,\delta=0.0001,\,\chi=0.002,\,  \alpha_1=1.2,\, \text{and}\, \beta_1=0.1$.}
\label{fig8}
\end{figure} 
\begin{figure*}[t]
\centering
\includegraphics[width=18.5cm,height=5cm]{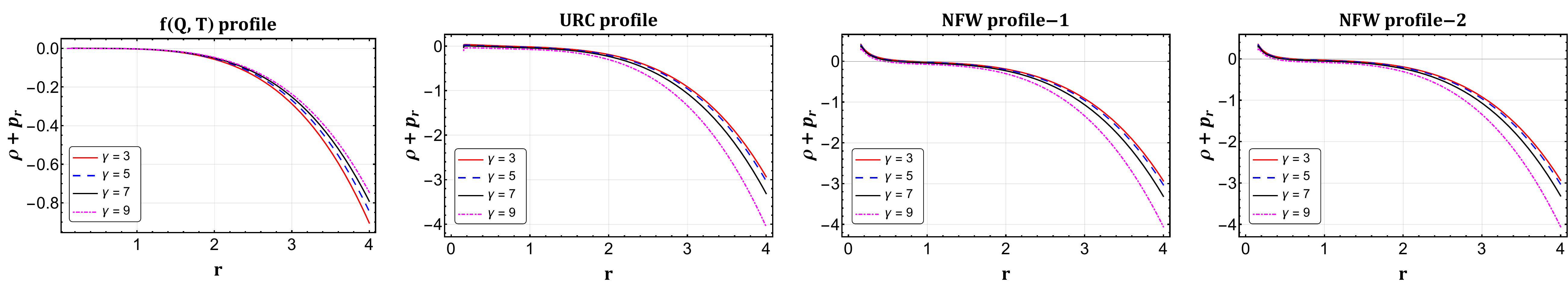}
\caption{Profile shows the behavior of NEC for radial pressure $\rho+p_r$ with respect to $r$ for different values of $\gamma$. We consider $r_0=1,\,\delta=0.0001,\,  \chi=0.002,\,\alpha_1=1.2,\, \beta_1=0.1,\,\rho_0=0.1,\,\text{and}\,r_s=1$.}
\label{fig9}
\end{figure*} 
\begin{figure*}[t]
\centering
\includegraphics[width=18.5cm,height=5cm]{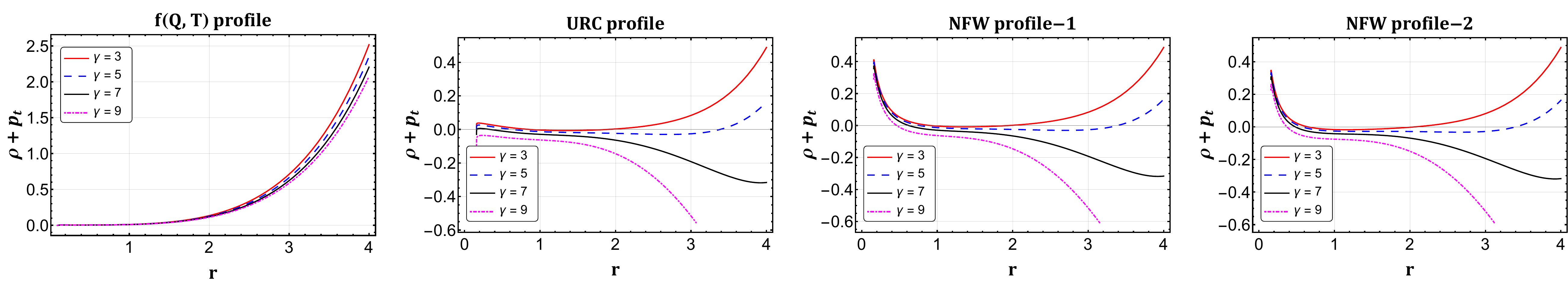}
\caption{Profile shows the behavior of NEC for tangential pressure $\rho+p_t$ with respect to $r$ for different values of $\gamma$. We consider $r_0=1,\,\delta=0.0001,\,  \chi=0.002,\, \alpha_1=1.2,\, \beta_1=0.1,\,\rho_0=0.1,\,\text{and}\,r_s=1$.}
\label{fig10}
\end{figure*} 
\begin{figure*}[t]
\centering
\includegraphics[width=18.5cm,height=5cm]{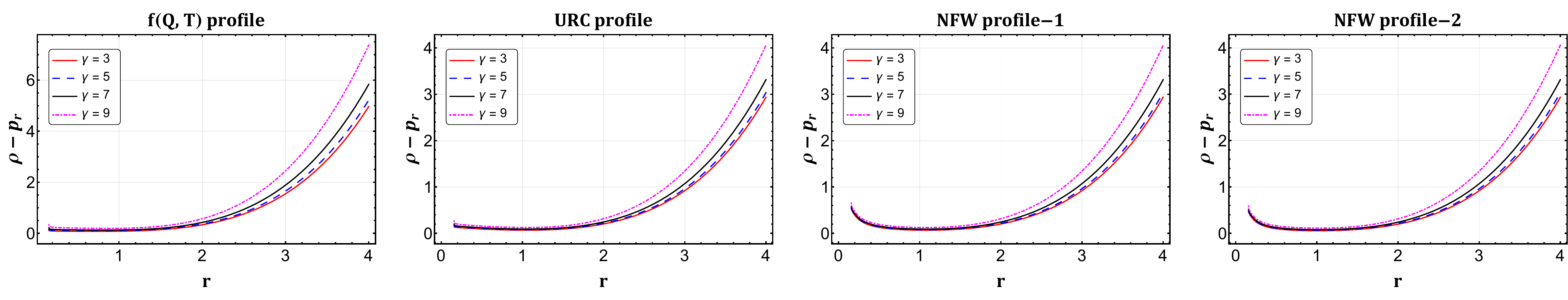}
\caption{Profile shows the behavior of DEC for radial pressure $\rho-p_r$ with respect to $r$ for different values of $\gamma$. We consider $r_0=1,\,\delta=0.0001,\, \chi=0.002,\, \alpha_1=1.2,\, \beta_1=0.1,\,\rho_0=0.1,\,\text{and}\,r_s=1$.}
\label{fig11}
\end{figure*} 
\begin{figure*}[t]
\centering
\includegraphics[width=18.5cm,height=5cm]{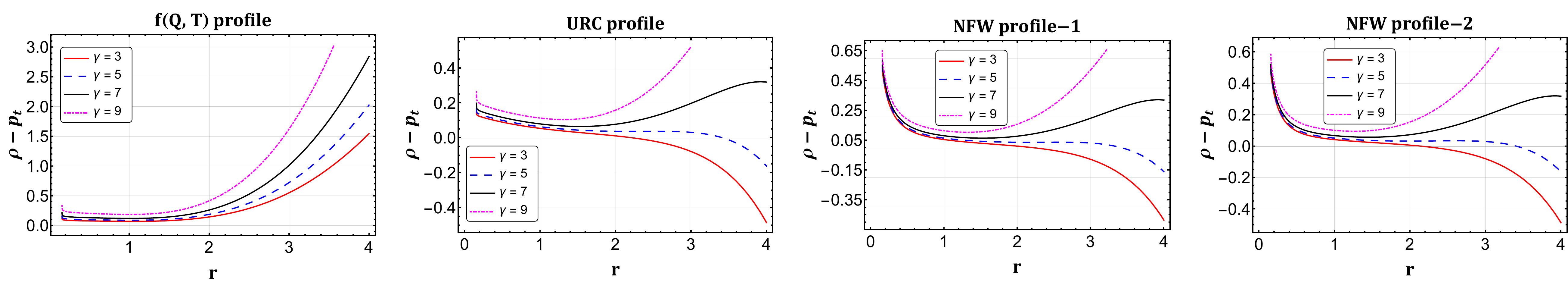}
\caption{Profile shows the behavior of DEC for tangential pressure $\rho-p_t$ with respect to $r$ for different values of $\gamma$. We consider $r_0=1,\,\delta=0.0001,\,  \chi=0.002,\, \alpha_1=1.2,\, \beta_1=0.1,\,\rho_0=0.1,\,\text{and}\,r_s=1$.}
\label{fig12}
\end{figure*} 
\begin{figure*}[t]
\centering
\includegraphics[width=18.5cm,height=5cm]{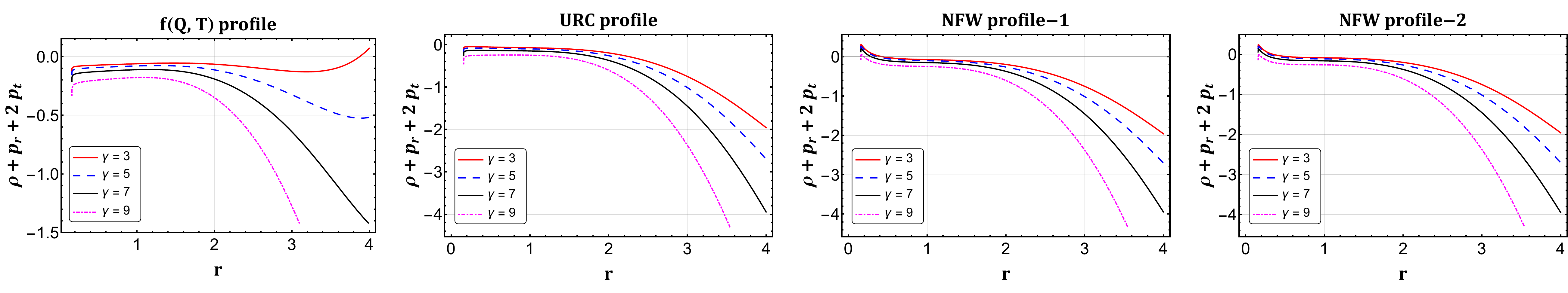}
\caption{Profile shows the behavior of SEC $\rho+p_r+2p_t$ with respect to $r$ for different values of $\gamma$. We consider $r_0=1,\,\delta=0.0001,\, \chi=0.002,\, \alpha_1=1.2,\, \beta_1=0.1,\,\rho_0=0.1,\,\text{and}\,r_s=1$.}
\label{fig13}
\end{figure*}

Now inserting the above shape function \eqref{46} with redshift function \eqref{44} into Eqs. (\ref{37}-\ref{39}) and presented the graphs for energy conditions in Figs. \ref{fig8}-\ref{fig13}.

We present the plotted graphs illustrating the energy conditions in Figures \ref{fig8}-\ref{fig13}. Fig. \ref{fig8} represents the graph depicting the relationship between energy density and radial coordinate $r$, demonstrating a positively increasing behavior throughout space-time, dependent on the model parameter $\gamma$. This behavior suggests that as $\gamma$ increases, the energy density also increases, indicating a potentially significant role of $\gamma$ in the energy distribution. Fig. \ref{fig9} displays the negative behavior of the radial NEC for various values of $\gamma$, resulting in a violation of the NEC. Furthermore, Fig. \ref{fig10} illustrates the positive behavior of tangential NEC for the $f(Q,T)$ profile, while the remaining profiles exhibit a negative behavior for different values of $\gamma$. This observation suggests that the $f(Q, T)$ profile possesses unique characteristics that allow for a positive contribution to the sum of energy density and tangential pressure. On the other hand, the negative behavior exhibited by the other profiles implies a dominance of negative pressure or energy density. Consequently, the NEC is violated for all profiles, indicating that exotic matter may sustain wormhole solutions in the context of non-metricity-based gravity, similar to GR. Additionally, we observe that the radial and tangential pressure satisfy the DEC for different $\gamma$ values, as shown in Figs. \ref{fig11} and \ref{fig12} respectively. Moreover, Fig. \ref{fig13} demonstrates the violation of the SEC for various $\gamma$ values. For more detailed information and quantitative analysis, one can refer to Table \ref{Table 2}.

\begin{table*}
\begin{center}
\begin{tabular}{|c|c|c|c|c|}
\hline
Energy conditions  & $f(Q, T)$ profile & URC profile & NFW profile-1 & NFW profile-2  \\ \hline
$\rho$ & $satisfied$ & $satisfied$ & $satisfied$ & $satisfied$ \\ 
\hline
$\rho + p_r$  & $violated$ & $violated$ & $violated$ & $violated$ \\ 
\hline
$\rho + p_t$ & $satisfied$ & $violated$ & $violated$ & $violated$ \\ 
\hline
$\rho - p_r$ & $satisfied$ & $satisfied$ & $satisfied$ & $satisfied$ \\ 
\hline
$\rho - p_t$ & $satisfied$ & $satisfied$ & $satisfied$ &  $satisfied$ \\ 
\hline
$\rho + p_r + 2 p_t$ & $violated$ & $violated$ & $violated$ & $violated$ \\ 
\hline
\end{tabular}
\caption{Description of Non-Linear Model}
\label{Table 2}
\end{center}
\end{table*}
\section{Final remarks}
Over the past few decades, the scientific investigations of wormhole geometry triggered much excitement among researchers. As a consequence, the URC and NFW DM density profiles supported the possibility of wormholes in the galactic halo region. In this paper, we have investigated the wormhole geometry affected by DM galactic halo profiles, namely URC and cold DM halo with NFW model-I and NFW model-II, under the background of non-metricity based modified $f(Q,T)$ gravity. It is known that it would be difficult to obtain the exact solutions with the arbitrary $f(Q,T)$ function; hence to overcome this issue, we have considered two specific models, such as linear $f(Q,T)= \alpha Q + \beta T$ and non-linear $f(Q,T)=\alpha_1 +\beta_1  \log (Q)+\gamma  T$ models (where $\alpha$, $\beta$, $\alpha_1$, $\beta_1$, and $\gamma$ are model parameters). The key findings of the current investigation are exclusively addressed below:\\
\indent First, we discussed the linear $f(Q, T)= \alpha Q + \beta T$ model under the DM halo profiles. For this model, we have obtained the shape functions by comparing the energy density of DM halo profiles with the energy density of $f(Q,T)$ gravity. We have investigated essential properties such as the flare-out condition of the obtained shape functions under asymptotic background. It is important to note here that the parameters of the involved model play a crucial influence in analyzing the shape of the wormholes. We noticed that the flare-out condition is satisfied near the throat; however, for very values of model parameter $\beta$, this condition may violate at the throat. Moreover, we checked the NEC under DM galactic halo models in the vicinity of the throat. It was observed that NEC is violated at the throat within the specific range of $\beta$, should be $-25.1327<\beta <19.2283$ (URC model), $-25.1327<\beta <19.2283$ (NFW model-I), and $-25.1327<\beta <26.6957$ (NFW model-II) corresponding to fixed values of parameters $\alpha=1.2,\,\rho_0=0.1,\, r_s=1\, \text{and} \, r_0 = 1$. Interestingly, we noticed that the contribution of the violation of NEC becomes higher if we increase the value of the model parameter $\beta$ within the mentioned range. Further, we noticed that DEC is satisfied, whereas SEC is disrespected at the throat of the wormhole. Moreover, we have discussed the energy conditions in detail in section- \ref{Diss} as well as summarized in Table- \ref{Table 1}.\\
\indent In the last part of this paper, we have investigated the wormhole solutions for the non-linear $f(Q, T)=\alpha_1 +\beta_1  \log (Q)+\gamma  T$ under the DM halo profiles. We have obtained wormhole solutions using the Karmarkar condition with embedded class-1 space-time. For the embedded shape function, we have studied the energy conditions under the effect of URC and NFW dark matter galactic halo profiles. It was observed that NEC for radial pressure is violated whereas, for tangential pressure, it is satisfied under $f(Q, T)$ profile while NEC for both pressures is violated under remaining DM halo profiles (see Figs. \ref{fig9} and \ref{fig10}). Also, the remaining energy conditions have been observed, such as DEC being satisfied for both pressures and SEC being violated under all the profiles near the throat. By constraining the model's parameter values and ranges, it is possible to demonstrate the violation of energy conditions, which supports the existence of exotic types of matter. Such a substance could make it possible for wormholes to travel through embedded space-time in the background. Moreover, the energy conditions for all the DM halo profiles and the $f(Q,T)$ profile are summarized in Table \ref{Table 2}. Thus, embedded wormhole solutions are physically acceptable within the DM galactic halo profiles.\\
\indent Recently, Rahaman et al. \cite{Rahaman11} studied wormhole solutions supported by dark matter and global monopole charge with semiclassical effects in the context of general relativity. They claimed that for fixed values of semiclassical effects, the monopole charge could be responsible for the violation of NEC. In \cite{Mehedi}, the authors investigated wormhole geometry with global monopole charge under some halo DM profiles. They argued that the monopole charge could minimize the violation of NEC. In this article, we have presented wormhole solutions without monopole charge under DM halo profiles in the background of generalized symmetric teleparallel gravity. Here, we can see the crucial effect of model parameters on wormhole solutions, which are mentioned above in the presented solutions. Further, it would be interesting to study wormhole solutions with global monopole charge under galactic halo profiles in teleparallel as well as symmetric teleparallel theories of gravity so that one could see the effect of monopole charge in modified theories of gravity.

\section*{Data Availability Statement}
 There are no new data associated with this article.

\acknowledgments MT acknowledges University Grants Commission (UGC), New Delhi, India, for awarding National Fellowship for Scheduled Caste Students (UGC-Ref. No.: 201610123801). ZH acknowledges the Department of Science and Technology (DST), Government of India, New Delhi, for awarding a Senior Research Fellowship (File No. DST/INSPIRE Fellowship/2019/IF190911). PKS acknowledges National Board for Higher Mathematics (NBHM) under the Department of Atomic Energy (DAE), Govt. of India, for financial support to carry out the Research project No.: 02011/3/2022 NBHM(RP)/R\&D II/2152 Dt.14.02.2022.

\end{document}